\documentclass{aa}  

\usepackage{graphicx}
\usepackage[varg]{txfonts}
\bibpunct{(}{)}{;}{a}{}{,}

\begin{document}

   \title{KIC\,9533489: a genuine $\gamma$ Doradus\,--\,$\delta$ Scuti \textit{Kepler} hybrid pulsator with transit events}
   
   \authorrunning{Zs. Bogn\'ar et al.}

   \titlerunning{Investigation of KIC\,9533489}


   \author{Zs. Bogn\'ar\inst{1}
          \and
           P. Lampens\inst{2}
          \and
           Y. Fr\'emat\inst{2} 
          \and
	   J. Southworth\inst{3}
	  \and
	   \'A. S\'odor\inst{1,2}
	  \and
	   P. De Cat\inst{2}
	  \and
	   H. T. Isaacson\inst{4}
	  \and
	   G. W. Marcy\inst{4}
	  \and
	   D. R. Ciardi\inst{5}
	  \and
	   R. L. Gilliland\inst{6}
	  \and
	  P. Mart\'{i}n-Fern\'andez\inst{7}
          }

   \institute{
   Konkoly Observatory, MTA CSFK, Konkoly Thege M. u. 15-17, H--1121 Budapest, Hungary\\
   \email{bognar@konkoly.hu}
   \and
   Koninklijke Sterrenwacht van Belgi\"e, Ringlaan 3, 1180 Brussel, Belgium
   \and
   Astrophysics Group, Keele University, Staffordshire ST5 5BG, UK
   \and
   Astronomy Department, University of California, Berkeley, CA 94720, USA
   \and
   NASA Exoplanet Science Institute/Caltech, Pasadena, CA 91125, USA
   \and
   Center for Exoplanets and Habitable Worlds, The Pennsylvania State University, University Park, PA 16802, USA
   \and
   Departamento de F\'isica Te\'orica y del Cosmos, Universidad de Granada, Campus de Fuentenueva, E-18071 Granada, Spain
              }

   \date{Received  ; accepted }

 
  \abstract
   {Several hundred candidate hybrid pulsators of type A--F have been identified from space-based observations.
   Their large number allows both statistical analyses and detailed investigations 
   of individual stars. This offers the opportunity to study the full interior of the genuine hybrids, in which both 
   low-radial-order \textit{p}-
   and high-order \textit{g}-modes are self-excited at the same time. However, a few other physical processes can also be responsible 
   for the observed hybrid nature, related to binarity or to surface inhomogeneities. The finding that most 
   $\delta$\,Scuti stars also show long-period light variations represents a real challenge for theory.}
   {We aim at determining the pulsation frequencies of \mbox{KIC\,9533489}, to search for regular patterns and spacings among them, and 
   to investigate the stability of the frequencies and the amplitudes.
   An additional goal is to study the serendipitously detected transit events: is KIC\,9533489 the host star? 
   What are the limitations on the physical parameters of the involved bodies?}
   {Fourier analysis of all the available \textit{Kepler} light curves. Investigation of the frequency and period spacings. 
   Determination of the stellar physical parameters from spectroscopic observations. Modelling of the transit events.}
   {The Fourier analysis of the \textit{Kepler} light curves revealed 55 significant frequencies 
   clustered into two groups, 
   which are separated by a gap between 15 and 27\,d$^{-1}$. 
   The light variations are dominated by the beating of two dominant frequencies located at around 4\,d$^{-1}$. The 
   amplitudes of these two frequencies show a monotonic long-term trend.
   The frequency spacing analysis revealed two possibilities: the pulsator is either a highly inclined moderate 
   rotator ($v\approx 70$\,km\,s$^{-1}$, $i > 70^\circ$) or a fast rotator ($v\approx 200$\,km\,s$^{-1}$) with 
   $i\approx20^\circ$.
   The transit analysis disclosed that the transit 
   events which occur with a $\approx197$\,d period may be caused by a $1.6\,R_{\rm Jup}$ body orbiting a fainter 
   star, which would be spatially coincident with KIC\,9533489.}
   {}


   \maketitle
%

\section{Introduction}

Recently, ultra-precise photometric data from three space missions revolutionized the field of variable star studies: 
the Canadian \textit{MOST} satellite \citep{2003PASP..115.1023W}, the French-led \textit{CoRoT} mission 
\citep{2009A&A...506..411A} and the NASA mission \textit{Kepler} \citep{2010ApJ...713L..79K}. The almost continuous, 
high-precision single-band light curves allow the detection of many pulsation frequencies with amplitudes down to the micromagnitude 
level. All these missions provide an enormous amount of extremely high-quality data for analysis, enabling 
state-of-the-art asteroseismic studies for many years to come. On the other hand, we still need complementary ground-based data for 
an unambiguous interpretation of the light variations observed by space telescopes, e.g., to determine the physical parameters 
and modal content of the pulsators, or to identify the (mostly non-eclipsing) binary/multiple systems.

A large fraction of intermediate-mass (1.2--2.5\,$M_\odot$) A and F spectral type 
stars shows pulsations in the intersecting region of the main sequence and the classical instability 
strip in the Hertzsprung--Russell diagram (HRD). Among others, we find here $\delta$\,Scuti ($\delta$\,Sct) variables, 
pulsating in low-radial-order 
\textit{p}- and mixed \textit{p}-\textit{g}-modes with periods in the $0.01 - 0.3$\,d range, and $\gamma$\,Doradus 
($\gamma$\,Dor) pulsators with high-order \textit{g}-modes in the $0.3 - 3$\,d period range. The $\delta$\,Sct pulsations 
are excited by the $\kappa$ mechanism operating in the HeII partial ionization zone (see e.g. the review by \citealt{2000ASPC..210....3B}), 
while the $\gamma$\,Dor pulsations are driven by the convective blocking mechanism operating at the base of the convection zone
\citep{2000ApJ...542L..57G}. The instability regions of these two mechanisms overlap in the HRD, and 
\textit{hybrid $\delta$\,Sct\,--\,$\gamma$\,Dor pulsators} were first predicted \citep{2004A&A...414L..17D}, subsequently 
confirmed \citep{2005AJ....129.2026H}. 
We refer to these as \textit{hybrids} hereafter. The simultaneously excited \textit{p}- and \textit{g}-modes provide information
respectively on the envelope and the near core region of the star. 
Therefore, such targets are very important targets for asteroseismic investigations. However, it is difficult to find these 
hybrids with ground-based observations because the $\gamma$\,Dor pulsations have usually low amplitudes, and their typical 
frequencies, close to 1--2\,d$^{-1}$, are strongly affected by the daily aliasing, as well as by atmospheric effects and 
instrumental drifts.

Thanks to the recent space-based observations, we now have hundreds of stars classified as A--F-type \textit{hybrid candidates} 
(see e.g. \citealt{2011A&A...534A.125U, 2012AN....333.1048H}). This allows both statistical analyses and detailed 
investigations of individual stars. The most striking result of the preliminary analyses of the hybrid candidates 
is that they are not confined to the theoretically predicted intersecting region of the two pulsation types. Instead, they occur everywhere 
in the $\delta$\,Sct domain \citep{2010ApJ...713L.192G, 2011A&A...534A.125U}, and it seems that practically all $\delta$\,Sct 
stars show low-frequency light variations \citep{2014MNRAS.437.1476B}. In the case of the hotter $\delta$\,Sct pulsators 
($T_{\rm eff} > 7500$\,K), the presumed existence of hybrid pulsations challenges the theory of stellar pulsations, because we do not 
expect that the convective blocking mechanism can operate in the thin convective upper layer \citep{2014MNRAS.437.1476B}. Furthermore, most of 
the candidates show frequencies between about 5 and 10\,d$^{-1}$, even though this region has formerly been considered as a 
`frequency gap', at least for low spherical degree ($l = 0-3$) modes (see Fig.~2 of \citealt{2010ApJ...713L.192G}).
This observational finding could be explained by the combined effects of frequency shifts of high-order \textit{g}-modes 
caused by the Coriolis force, detection of high-spherical-degree modes, and rotational splitting \citep{2013MNRAS.429.2500B}. 

What mechanisms are responsible for the simultaneous occurrence of low- and high-frequency variations revealed in so many A--F-type stars? 
Different possible scenarios have to be considered: 

\begin{enumerate}
 \item \textit{Genuine $\delta$\,Sct--$\gamma$\,Dor hybrids}: both $\delta$\,Sct and $\gamma$\,Dor-type pulsations 
 are self-excited in these objects. About a dozen such confirmed objects have been investigated in detail, but there must be many 
 more genuine hybrids among the known candidates.
 \item \textit{Eclipsing or ellipsoidal binaries}: the object might be a $\delta$\,Sct pulsator in an eclipsing or 
 ellipsoidal binary system. The long periodicity in the light variations corresponds to the orbital period or half of it,
 depending on the case. The physical 
 parameters of the binary components can be derived independently from any theoretical (asteroseismic) assumption. 
 These parameters can be used as constraints for subsequent asteroseismic investigations.
 \item \textit{Tidal excitation}: a $\delta$\,Sct star in a close, eccentric binary, where the external tidal forces excite 
 \textit{g}-mode pulsations \citep{2002A&A...384..441W}. \object{HD 209295} is an example \citep{2002MNRAS.333..262H}.
 \item \textit{Stellar rotation coupled with surface inhomogeneities}: these can also cause detectable signals in the 
 low-frequency region. Low temperature spots may be an obvious explanation, but mechanisms generating 
 such features are not expected to operate in normal (non-Ap, non-Am) A-type stars, since their convective layers are thin and thus likely 
 have weak magnetic fields. Nevertheless, statistical investigation of \textit{Kepler} A-type stars 
 \citep{2011MNRAS.415.1691B} shows that the low-frequency variability could be caused by rotational modulation in connection 
 with starspots or other corotating structures. 
 Low temperature spots cause characteristic light-curve variations. 
 One low-frequency signal and several of its harmonics are expected in the Fourier spectrum. However, finite starspot 
 lifetimes and/or differential rotation combined with spot migration result in a number of significant peaks around the 
 rotational frequency on longer time scales.
\end{enumerate}

 However, in the case of \object{KIC\,5988140}, none of the above mentioned simple scenarios could explain the observations 
 in a fully satisfactory way \citep{2013A&A...549A.104L}. Indeed, in this case, the scenario of binarity has been discarded 
 on the basis of the double-wave pattern of the radial velocity curve mimicking the pattern of the light curve (after prewhitening 
 of the higher frequencies). The other two remaining scenarios explaining the low frequencies in terms of $g$-mode pulsation or 
 rotational modulation are also problematic since the predicted light-to-velocity amplitude ratio based on two simple spotted 
 surface models is 40 times larger than observed.

\object{KIC\,9533489} is another \textit{Kepler} hybrid candidate investigated by \citet{2011A&A...534A.125U}, and 
which turns out as an exotic case: in addition to its confirmed hybrid character, we also detected a few transit events 
in the \textit{Kepler} light curve. In the following sections, we present the results of the frequency analyses and derived 
(physical and pulsational) properties of this object, as well as a possible model for the transit events detected in the light curve.

\section[]{The \textit{Kepler} data}

KIC\,9533489 ($Kp=12.96$\,mag, $\alpha_{2000}=19^{\mathrm h}38^{\mathrm m}41.7^{\mathrm s}$, 
$\delta_{2000}=+46^{\mathrm d}07^{\mathrm m}21.6^{\mathrm s}$) was observed both in long-cadence (LC) and short-cadence
(SC) mode. The LC data have a sampling time of 29.4 minutes \citep{2010ApJ...713L.120J}, while the SC observations provide 
a brightness measurement every 58.85 seconds \citep{2010ApJ...713L.160G}. We used the `raw' flux data of KIC\,9533489 
available from the \textit{Kepler} Asteroseismic Science Operations Center (KASOC) database. We analysed the 
LC data of 17 quarters of \textit{Kepler} observations (Q1--Q17, 13th May 2009 -- 11th May 2013) and the 
SC data of 3 quarters (Q1, Q5.1--3, Q6.1--3, 13th May 2009 -- 22nd September 2010). The contamination level 
of the measurements is estimated to be only 1.6 per cent.

The strong long-term instrumental trends present in the light curves were removed before analysis. First, we 
divided the time strings into segments. These segments contain gaps no longer than 0.1--0.5\,d. Then, we separately 
fitted and subtracted cubic splines from each segment. We used one knot point for every 1000 (SC data) or 100--200 points 
(LC data) to define the splines. We omitted the obvious outliers and the data points affected by the transit 
events. The resulting SC and LC light curves consist of 307\,950 and 62\,418 data points, spanning 498 (with a 
large gap) and 1450 days, respectively.

\section[]{Spectroscopy}
\label{sect:spec}

For an unambiguous interpretation of the results of the light-curve analysis, the atmospheric stellar parameters, i.e. the effective 
temperature ($T_{\mathrm{eff}}$), surface gravity ($\mathrm{log}\,g$) and projected rotational velocity ($v\,\mathrm{sin}\,i$), 
must be known. Therefore, we observed KIC\,9533489 in the 2013 and 2014 observing seasons with the High Efficiency and 
Resolution Mercator Echelle Spectrograph (\textsc{hermes}, \citealt{2011A&A...526A..69R}) attached to the 1.2-m Mercator 
Telescope, with the FIbre-fed Echelle Spectrograph (\textsc{fies}) at the 2.5-m Nordic Optical Telescope (NOT), and also with the 
\textsc{integral} instrument \citep{1998SPIE.3355..821A} using the 4.2-m William Herschel Telescope (WHT). All instruments are 
located at the Roque de los Muchachos Observatory (ORM, La Palma, Spain). We also acquired data with the High Resolution Echelle Spectrometer 
(\textsc{hires}, \citealt{1994SPIE.2198..362V}) attached to the 10-m Keck\,I telescope (W.~M. Keck Observatory, Hawaii, U.S.).
Table~\ref{tabl:logobs} shows the journal of the spectroscopic observations.

\begin{table*}
\centering
\caption{\label{tabl:logobs} Log of spectroscopic observations of KIC\,9533489. 
}
\begin{tabular}{llcrcrcr}
\hline\hline
 \multicolumn{1}{c}{Telescope} & \multicolumn{1}{c}{Instrument} & \multicolumn{1}{c}{Wavelength range} 
 & \multicolumn{1}{c}{Resolution} & \multicolumn{1}{c}{Obs. Time} & \multicolumn{1}{c}{Exp. time} & Obs. & S/N\\
  &	       & \multicolumn{1}{c}{(nm)}		  & & \multicolumn{1}{c}{(JD)}	& \multicolumn{1}{c}{(s)} & \# & \\
\hline
1.2-m Mercator & \textsc{hermes} & 380--900 & 85\,000 & 2\,456\,514.4 & 1800 & 3 & 20\\
2.5-m NOT & \textsc{fies} & 370--730 & 45\,000 & 2\,456\,574.4 & 1800 & 3 & 55\\
4.2-m WHT & \textsc{integral} & 570--700 & 2\,750 & 2\,456\,783.7 & 600 & 1 & 70\\
10-m Keck\,I & \textsc{hires} & 386--678 &  67\,000 & 2\,456\,845.0 & 1132 & 1 & 100\\
\hline
\end{tabular}
\tablefoot{S/N is the signal-to-noise ratio
per sample of the mean spectrum computed in the $\lambda = 496 - 564$\,nm wavelength range 
(\textsc{hermes}, \textsc{fies} and \textsc{hires} spectra), or at $\lambda \sim 570$\,nm (\textsc{integral} spectrum),
and by adopting a procedure described in \citet{2007STECF..42....4S}. The \textsc{hires} spectrum has gaps
in the 477--498 and 628--655\,nm domains.
}
\end{table*}

We obtained the reduced spectra from the dedicated pipelines of the \textsc{hermes}, 
\textsc{fies} and \textsc{hires} instruments, while the single-order WHT spectrum centered 
on H$\alpha$ was reduced using \textsc{iraf}\footnote{\textsc{iraf} is distributed 
by the National Optical Astronomy Observatories, which are operated by the 
Association of Universities for Research in Astronomy, 
Inc., under cooperative agreement with the National Science Foundation.} routines.
All spectra were corrected for barycentric motion, then combined 
into a single averaged spectrum per instrument.

Though the signal-to-noise ratio (S/N) of the averaged \textsc{hermes} spectrum is 
low \mbox{(S/N = 20} per wavelength bin at $\lambda \sim 500$\,nm), calculating a cross-correlation 
function with an F0-star synthetic spectrum,
excluding regions of Balmer and telluric lines, 
we were able to derive the radial velocity (RV) for that epoch. We performed the same procedure
on the \textsc{fies} and \textsc{hires} spectra, too. 
We found that the RV measurements are 
consistent with each other within the errors, pointing to a value of $6.5\pm1.5$\,km\,s$^{-1}$. 
That is, the star showed no sign 
of RV change over a period of $\sim330$ days, and therefore cannot be considered
as a multiple system.

Because it has a sufficiently high S/N and that it covers a wide wavelength domain 
that contains all the majour Balmer lines, we started to derive the star's
astrophysical parameters on the \textsc{fies} averaged spectrum.
We estimated $v\,\mathrm{sin}\,i$ by adopting the Fourier transform (FT) approach
(for an overview, see \citealt{1992oasp.book.....G, 2005MSAIS...8..124R}, and references therein). 
To increase the S/N and to limit the impact of blends, the FT was performed on an average line 
profile obtained through cross-correlation of the spectrum with an F0 mask from which we excluded
the strongest lines. The result shows that the star is probably a moderate rotator 
(see Table~\ref{table:params}). 

We estimated $T_{\mathrm{eff}}$ by fitting the H$\alpha$, H$\beta$, H$\gamma$, 
and H$\delta$ line profiles of the \textsc{fies} spectrum. 
The continuum normalisation of the spectrum was realised in three
steps. A first normalisation was done with the \textsc{iraf}/\textsc{continuum} task
to perform a very preliminary determination of $T_{\mathrm{eff}}$ and $\mathrm{log}\,g$ in the
400--450\,nm range (metallicity was kept Solar and microturbulence was fixed
to 2\,km\,s$^{-1}$). The fit was realised by using the \textsc{girfit} program based
on the minuit minimization package and by interpolating in a grid of
synthetic spectra. These synthetic spectra and corresponding LTE model atmospheres
were computed with the \textsc{synspec} (\citealt{1995ApJ...439..875H} and references
therein), and with \textsc{atlas9} (\citealt{2004astro.ph..5087C} and references therein)
computer codes, respectively. They were then convolved
with  the rotational profile and with a Gaussian instrument profile to account for the resolution 
of the spectrograph. The observations were then divided by the closest, 
not normalised, synthetic spectrum. 

After being sigma clipped to remove most mismatch features,
the resulting curve was divided into 2 to 4 smaller regions which were smoothened
with polynomials of degrees 5 to 10. The observed spectrum was divided
by the resulting function (see Fig.~1 of \citet{2007A&A...471..675F} for
an illustration of the procedure) then normalised with the \textsc{iraf}/\textsc{continuum}
task and a low order spline. Regions around H$\alpha$, H$\beta$, H$\gamma$,
and H$\delta$ were then isolated and fitted separately. Because below 8000\,K,
hydrogen lines are only sensitive to effective temperature, all other
parameters were kept fixed and we tried as much as possible to exclude from the
fit most metal lines. Each time, 3 to 5 different starting points were chosen
between 6000 and 8000\,K, which resulted in a median value and
an inter-quantile dispersion of $7350\pm200$\,K.

The same procedure was then applied on the other spectra, except that fewer
hydrogen lines were available. We list the values we obtained in 
Table~\ref{tabl:teffs}, where the last column details the hydrogen lines
that we considered. For the \textsc{hermes} and \textsc{fies} spectra the
error bars were deduced from the scatter of our different determinations
and trials, while for the \textsc{integral} and \textsc{hires} determinations
we adopted the numerical value provided by the {\sc minuit} package when
it computes the covariance matrix. The final $T_{\mathrm{eff}}$ value
is obtained by combining these different estimated into a rounded
mean which error is provided by the largest deviation (see Table \ref{table:params}).

\begin{table}
\caption{\label{tabl:teffs} Effective temperature values derived from the fit
of the hydrogen line profiles observed with different instruments.}
\begin{tabular}{lll}
\hline\hline
 \multicolumn{1}{c}{Instrument} & \multicolumn{1}{c}{$T_{\mathrm{eff}}$ (K)} 
 & Line profiles considered \\
\hline
\textsc{hermes} & 7250 $\pm$ 150 & H$\alpha$, H$\beta$, H$\gamma$ \\
\textsc{fies} & 7350 $\pm$ 200 & H$\alpha$, H$\beta$, H$\gamma$, H$\delta$ \\
\textsc{integral} & 7200 $\pm$ 20 & H$\alpha$ \\
\textsc{hires} & 7190 $\pm$ 70 & H$\delta$, H$\gamma$\\
\hline
\end{tabular}
\end{table}

Since our initial scope was not to perform a detailed chemical abundance analysis,
we derived the metal content by scaling the abundance of all elements heavier
than boron by the same factor which we call here after `metallicity'.
For this purpose we studied the region between 496 and 564\,nm. To identify the features 
that are the most sensitive to surface gravity, we considered two synthetic spectra 
of same effective temperature (7200\,K) and $v\,\mathrm{sin}\,i$ but having different 
extreme $\mathrm{log}\,g$ values (3.5 and 5.0). 
The difference between the two spectra exceeds the limits given by the S/N of the observed 
spectra only in the region around the Mg\,I triplet (515 to 520\,nm).
We then constructed a first regular grid of synthetic spectra
for different values of the microturbulent velocity and metallicity.
The $\chi_2$ value between the observations and the spectra of each ($v_\mathrm{mic}$, 
metallicity) node was computed from 496 to 564\,nm, excluding the 515--520\,nm
part. Finally, a Levenberg-Marquardt minimization algorithm was applied
around the minimum and a covariance matrix was computed to estimate the error
bars by interpolating in a second grid of spectra having smaller steps in metallicity and
microturbulence. $\mathrm{Log}\,g$ was then derived by fixing the microturbulence and
metallicity to the best values found in the previous step, and by applying the
Levenberg-Marquardt algorithm to the region between 515--520\,nm. Because, the
final $\mathrm{log}\,g$ value did not have any significant impact on the agreement between
synthetic spectrum and observations in the other considered wavelength domains 
we stopped the process. 

\begin{figure}
\begin{center}
\includegraphics[width=8.7cm]{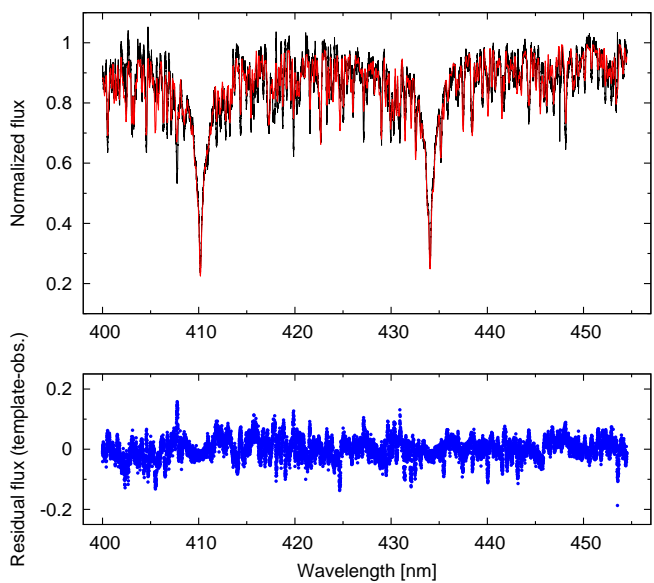}
\end{center}
\caption{Observed \textsc{hires} spectrum (in black) and synthetic model (in red) of an F0-type star with 
$v\,\mathrm{sin}\,i = 70$\,km\,s$^{-1}$ over the wavelength range 400--455\,nm. The residual spectrum is drawn in blue.}
\label{fig:specfit}
\end{figure}

Part of the \textsc{hires} spectrum is shown in Fig.~\ref{fig:specfit} where it
is compared to the synthetic spectrum that results from our atmospheric parameter determination procedure.
The residuals between the two spectra are also plotted
in the same figure. These residuals are significant compared to the S/N.
We checked therefore whether we could find any trace of an additional contribution
in the spectrum as suggested by the transit modelling (see Sect.~\ref{sect:transit}).

First, we computed the cross-correlation of the \textsc{fies} spectrum with a G5
spectral template. We did not find any evidence of absorption lines due to a cooler
companion. This could mean that the presumed companion's contribution to the total
light is too low to be detected in the spectrum and/or that its spectral lines
are too broadened (i.e. due to rotation or/and macroturbulence). A light
factor corresponding to $\Delta\mathrm{m} = 2.25$\,mag might explain such non-detection.

We next computed an (F0+G5) composite model spectrum and looked at the variance between
this model and the \textsc{hires} spectrum in the wavelength range 500--560\,nm for
a wide range of possible radial velocities. The G5-synthetic spectrum was computed assuming
$v\,\mathrm{sin}\,i = 10$\,km\,s$^{-1}$. The minimum variance was found
around $31$\,km\,s$^{-1}$, i.e. different from the radial
velocity of the \textit{Kepler} star.
This second test might hint toward the existence
of an additional stellar object, but we remain unable to identify the lines of the
transit host directly in the spectrum.

\begin{figure}
\begin{center}
\includegraphics[width=8.9cm]{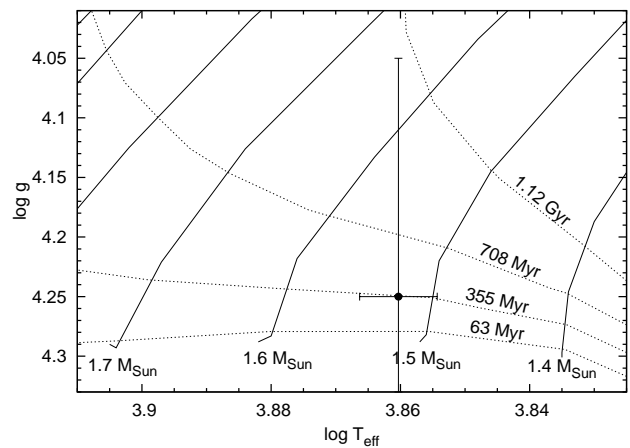}
\end{center}
\caption{Location of KIC\,9533489 on the surface gravity vs. effective temperature diagram.
Evolutionary tracks for 1.4, 1.5, 1.6, and 1.7\,$M_\odot$, as derived by \citet{2000A&AS..141..371G}, 
are plotted with solid lines. Dotted lines show 63\,Myr\,--1.12\,Gyr isochrones.}
\label{fig:hrd}
\end{figure}

Table~\ref{table:params} summarizes the stellar parameters derived from the spectra. 
Knowing the values of $T_{\mathrm{eff}}$ and $\mathrm{log}\,g$, and assuming a solar metallicity, we estimated the mass, 
luminosity, radius and age of KIC\,9533489 (see also in Table~\ref{table:params}) applying the evolutionary tracks and 
isochrones from \citet{2000A&AS..141..371G}. 
The $T_{\mathrm{eff}}$ and $\mathrm{log}\,g$ values indicate that KIC\,9533489 is situated in the 
overlapping region of the $\delta$\,Sct and $\gamma$\,Dor observational instability strips (see e.g. fig.~10 of 
\citealt{2011A&A...534A.125U}). The age estimation suggests a young object, located near the 
zero-age main sequence (ZAMS), but the uncertainty is very large (Fig.~\ref{fig:hrd}).

\begin{table}
\centering
\caption{\label{table:params} Stellar parameters of KIC\,9533489, determined from spectroscopic observations 
(\textsc{`spectra'}) and evolutionary tracks and isochrones (\textsc{`models'}, \citealt{2000A&AS..141..371G}).}
\begin{tabular}{lrlr}
\hline\hline
 & \multicolumn{1}{c}{\textsc{spectra}} & & \multicolumn{1}{c}{\textsc{models}}\\
\hline
$T_{\mathrm{eff}}$ & $7250\pm100$\,K & $M_*$ & $1.52^{+0.15}_{-0.05}\,M_\odot$\\[1mm]
$\mathrm{log}\,g$ & $4.25\pm0.2$\,dex & $L_*$ & $5.8^{+2.1}_{-1.2}\,L_\odot$\\[1mm]
$v\,\mathrm{sin}\,i$ & $70\pm7$\,km\,s$^{-1}$ & $R_*$ & $1.53^{+0.35}_{-0.15}\,R_\odot$\\[1mm]
$v_\mathrm{mic}$ & $2.9\pm0.3$\,km\,s$^{-1}$ & Age & $350^{+750}_{-350}$\,Myr\\[1mm]
$[$M/H$]$ & $-0.05\pm0.2$ & $\overline{\rho_*}$ & $0.42^{+0.17}_{-0.21}\,\rho_\odot$ \\[1mm]
$V_\mathrm{rad}$ & $6.5\pm1.5$\,km\,s$^{-1}$ & & \\
\hline
\end{tabular}
\end{table}

\section{Adaptive optics observations}
\label{sect:ao}

We obtained near-infrared adaptive optics (AO) images of KIC\,9533489 in the
Br-$\gamma$ (2.157\,$\mu$m) filter using the NIRC2 imager behind the natural
guide star adaptive optics system at the Keck\,II telescope
on 04 September 2014.  Data were acquired in a three-point dither pattern to
avoid the lower left corner of the NIRC2 array which is noisier than the
other three quadrants.  The dither pattern was performed three times with
step sizes of 3$\arcsec$, 2.5$\arcsec$, and 2.0$\arcsec$ for a total of nine
frames; individual frames had an integration time of 30\,s.
The individual frames were flat-fielded, sky-subtracted (sky
frames were constructed from a median average of the 9 individual source
frames), and co-added.

The average AO-corrected seeing was $\mathrm{FWHM} =0.048\arcsec$; the pixel scale
of the NIRC2 camera is 10\,mas/pixel. $5\sigma$ sensitivity
limits were determined for concentric annuli surrounding the primary targets;  the
annuli were stepped in integer multiples of the FWHM with widths of 1 FWHM.
The sensitivity achieved reached $\Delta\mathrm{mag} \approx 8$\,mag at $0.2\arcsec$
separation from the central target star in the band of observation.  Using typical stellar
\textit{Kepler}-infrared colours \citep{2012ApJ...746..123H}, this value translates
into the \textit{Kepler} bandpass to be about $\Delta\mathrm{mag}
\approx 10-11$\,mag.

\begin{figure}
\begin{center}
\includegraphics[width=8.5cm]{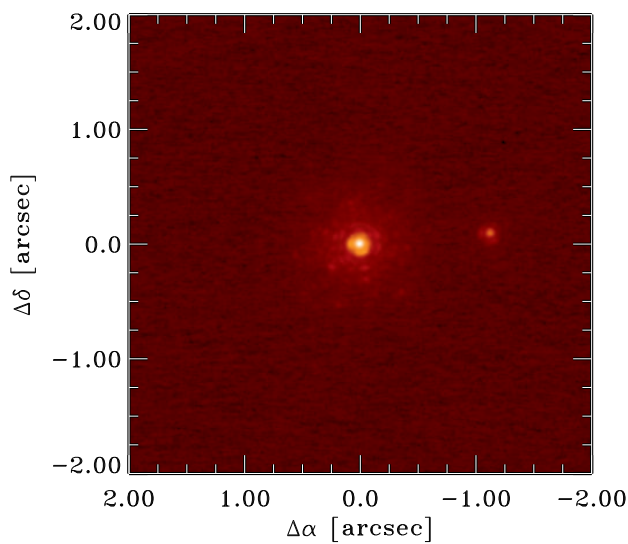}
\end{center}
\caption{Keck\,II adaptive optics image of KIC\,9533489.}
\label{fig:keckao}
\end{figure}

One source was detected approximately $1.1\arcsec$ to the west of the primary
target ($\Delta\alpha = 1.12 \pm 0.005\arcsec$; $\Delta\delta = 0.09\pm 0.005\arcsec$; Fig.~\ref{fig:keckao}).
The companion is $3.412 \pm 0.014$\,mag fainter than the target at $2.2\,\mu$m;
deblending the 2MASS $Ks$ photometry, we find the infrared brightnesses of the
target and the companion to be $Ks = 12.055 \pm 0.020$\,mag and $Ks = 15.467 \pm 0.024$\,mag.
No observations in other wavelengths were obtained, but based upon the typical
\textit{Kepler}-infrared colours \citep{2012ApJ...746..123H}, the companion has a \textit{Kepler} bandpass
magnitude of $Kp = 17.9 \pm 0.8$\,mag; thus, the companion is $\gtrsim 5$ magnitudes
fainter than the primary target and contributes less than 1\% to the \textit{Kepler}
light curve.

\section{Frequency analysis}

\begin{figure}
\begin{center}
\includegraphics[width=9.0cm]{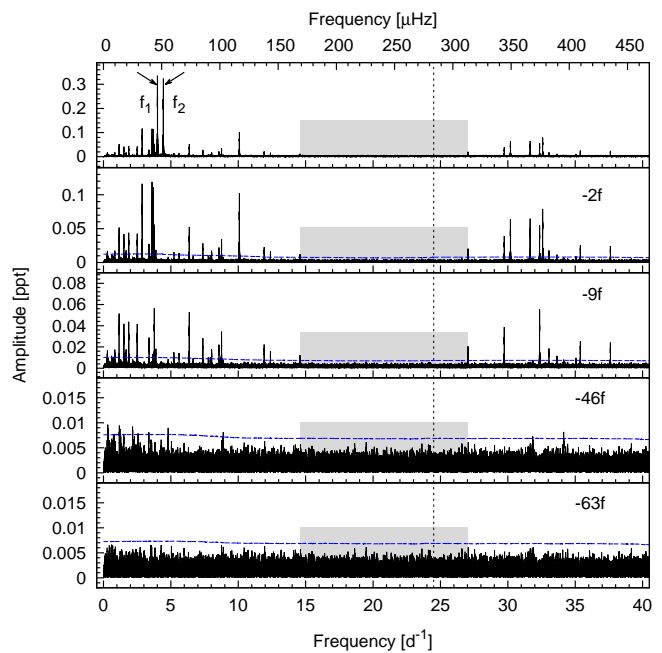}
\end{center}
\caption{Fourier transforms of the original SC light curve and of the light curve pre-whitened for 2, 9, 
46, and 63 frequencies. The blue dashed lines 
denote the $4\langle A \rangle$ significance level, where $\langle A \rangle$ is calculated as the moving 
average of radius 8\,d$^{-1}$ of the pre-whitened spectrum. The frequency gap between $\approx$15 and 
27\,d$^{-1}$ is indicated with gray band. The vertical dotted line denotes the Nyquist limit of the LC data.}
\label{fig:pw}
\end{figure}

\begin{figure*}
\begin{center}
\includegraphics[width=18cm]{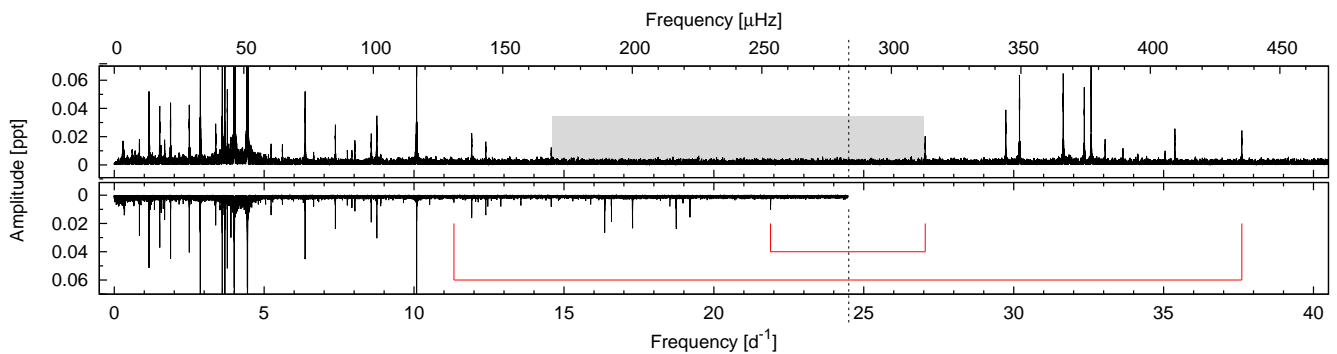}
\end{center}
\caption{Fourier transforms of the original SC (upper panel) and LC (bottom panel) light curve, respectively. 
The frequency gap between $\approx$15 and 27\,d$^{-1}$ found in the SC data is indicated with gray band. 
The vertical dotted line denotes the Nyquist limit of the LC data.
Red lines connect the lowest and the highest frequency peaks of the SC data with their 
Nyquist aliases detected in the LC Fourier transform.}
\label{fig:lcsc}
\end{figure*}

We performed the Fourier analysis of both the SC and LC data sets and compared the frequencies obtained.
The increased sampling of the SC data allows the direct determination of pulsation frequencies above the LC Nyquist limit 
($f_{\mathrm{LC,Nyq}}=24.5$\,d$^{-1}$, $f_{\mathrm{SC,Nyq}}=734$\,d$^{-1}$). The LC observations, having a longer 
time base, provide better frequency resolution and allow the investigation of the long-term frequency stability.
The Rayleigh frequency resolutions of the long- and short-cadence data sets are $f_{\mathrm{LC,R}}=0.0007$ and 
$f_{\mathrm{SC,R}}=0.002$\,d$^{-1}$, respectively.

We performed a standard successive prewhitening of the light curves for significant signals using \textsc{Period04} 
\citep{2005CoAst.146...53L} and \textsc{SigSpec} \citep{2007A&A...467.1353R}. The results were also checked with the photometry 
modules of \textsc{famias} \citep{2008CoAst.155...17Z}. The frequencies obtained were consistent within the errors 
using the different programs.
Fig.~\ref{fig:pw} presents the Fourier transforms of the original and the pre-whitened 
SC data sets. We stopped the process when none of the remaining peaks reached the threshold of S/N\,$\approx4$ 
\citep{1993A&A...271..482B}. 
Altogether, 63 frequencies were determined in this way. 
We also checked the spectral significances (sig) of the frequencies, as derived by \textsc{SigSpec}.
All of the peaks were found to be above the default threshold value for determining frequencies (5.46, see \citealt{2007A&A...467.1353R}).
The lowest amplitude peaks have sig$\sim$8 (cf. Table~\ref{table:freq}).
We checked which of these frequencies 
can also be found in the LC data. Below the LC Nyquist limit, we found all but eight frequencies. All of the eight 
missing frequencies are below 1\,d$^{-1}$. Both the instrumental effects in the original data (see e.g. 
\citealt{2012MNRAS.422..665M}) and the correction method we applied to remove these affect this low-frequency region. 
Therefore, we excluded these eight peaks from the list of accepted frequencies. Checking the remaining
significant frequencies using the $\textit{Kepler}$ Data Characteristics Handbook (KSCI-19040-004), none of them was found 
to be of instrumental origin.

Note that the frequencies detected in the SC data above the LC Nyquist limit, except for the three lowest amplitude ones, 
also appear in the LC data.
These can be identified as Nyquist aliases at $f_\mathrm{i,LC} = 2 f_\mathrm{LC,Nyq} - f_\mathrm{i,SC}$, split 
by \textit{Kepler}'s
orbital frequency \citep{2013MNRAS.430.2986M}. Fig.~\ref{fig:lcsc} presents the FT of the original SC and LC data, respectively.
We can detect both real pulsation frequencies and such Nyquist aliases above 11.3\,d$^{-1}$
in the Fourier transform of the LC light curve.

Table~\ref{table:freq} lists the 55 accepted frequencies and the differences between the frequency values derived from
the SC and LC data, too. They differ only in the fourth--sixth decimals. Note that 
the frequency difference (0.0056\,d$^{-1}$) of the closest peaks ($f_{\mathrm{16}}$ and $f_{\mathrm{51}}$) is 2.8 times 
larger than the Rayleigh frequency resolution of the SC data set, so all of the identified frequencies are well-resolved. 

\begin{figure}
\begin{center}
\includegraphics[width=8.5cm]{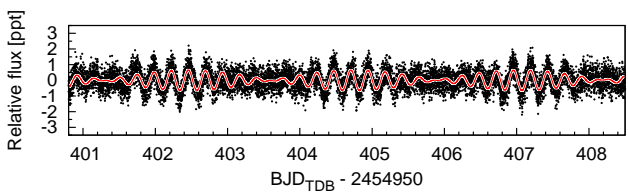}
\end{center}
\caption{A section of the SC light curve of KIC\,9533489 and the fit of the two dominant frequencies (red line). 
The plot exhibits the beating effect dominating the light variations.}
\label{fig:lc}
\end{figure}

Two closely spaced peaks at $\approx4$\,d$^{-1}$ 
($f_{\mathrm{1}}$ and $f_{\mathrm{2}}$) dominate the light variation of KIC\,9533489 (see Fig.~\ref{fig:lc}). 
Most of the peaks (42) are 
below 15\,d$^{-1}$, and 27 of these are below 5\,d$^{-1}$. The rest of the frequencies (13) are in the 27--38\,d$^{-1}$ 
range. There is a clear frequency gap between 15 and 27\,d$^{-1}$. We indicate this region with a gray band in 
Figs.~\ref{fig:pw} and \ref{fig:lcsc}. Such gaps play an important role in hybrid star studies, as these allow to distinguish between 
different mode-driving mechanisms.

We checked the frequencies for possible combination terms ($kf_{\mathrm{i}}$ or $pf_{\mathrm{i}} \pm qf_{\mathrm{j}}$). 
Considering test results on simulated light curves \citep{2012AN....333.1053P}, we 
restricted our search only to the second-order combinations, i.e. $|p|=1$ and $|q|=1$. We accepted a peak as a combination 
if the amplitudes of both parent 
frequencies were larger than that of their presumed combination term, and the difference between the observed and 
the predicted frequency was not larger than the Rayleigh resolution of the SC data ($0.002$\,d$^{-1}$). This way, we 
found one harmonic peak ($f_{\mathrm{49}} = 2f_{\mathrm{2}}$) and 11 other combination terms. Note that although $f_{\mathrm{30}}$ 
is close to $2f_{\mathrm{1}}$, it does not fulfil the above criterion. 
We indicate the combination terms in Table~\ref{table:freq}. Beside the combinations consisting of independent
parent frequencies, we list two additional cases in parentheses in Table~\ref{table:freq}, too. These are 
equal to two and four times the value of $f_{\mathrm{28}}$, respectively, which is itself a combination of two large-amplitude frequencies 
($f_{\mathrm{28}} = f_{\mathrm{2}}-f_{\mathrm{3}}$). It is not clear yet, whether this is a coincidence, or whether $f_{\mathrm{28}}$ 
(0.84\,d$^{-1}$) has a particular importance for some unknown reason.

\begin{table*}
\centering
\caption{\label{table:freq} Frequency content of the KIC\,9533489 \textit{Kepler} data.
}
\begin{tabular}{lrrrccrrr}
\hline\hline
 & \multicolumn{3}{c}{Frequency} & Ampl. & Phase & \multicolumn{1}{c}{S/N} & \multicolumn{1}{c}{sig} & Combinations\\
 & \multicolumn{1}{c}{[d$^{-1}$]} & \multicolumn{1}{r}{$\Delta f$\tablefootmark{a}} & 
 \multicolumn{1}{c}{[$\mu$Hz]} & [ppt] & [$2\pi$] & & & \\ 
\hline
$f_{\mathrm{1}}$ & 4.007089(4) & -0.04 & 46.37834 & 0.333 & 0.566 & 186.2 & 9967.3 & \\
$f_{\mathrm{2}}$ & 4.443236(4) & 0.10 & 51.42635 & 0.322 & 0.039 & 180.3 & 10647.4 & \\
$f_{\mathrm{3}}$ & 3.60269(1) & -0.23 & 41.6978 & 0.120 & 0.185 & 67.3 & 1747.5 & \\
$f_{\mathrm{4}}$ & 2.87088(1) & -0.33 & 33.2277 & 0.116 & 0.947 & 64.6 & 1693.5 & \\
$f_{\mathrm{5}}$ & 3.68992(1) & 0.12 & 42.7075 & 0.113 & 0.940 & 63.1 & 1646.5 & \\
$f_{\mathrm{6}}$ & 10.08357(1) & 0.27 & 116.7080 & 0.102 & 0.850 & 59.0 & 1385.9 & \\
$f_{\mathrm{7}}$ & 32.58125(2) &  & 377.0979 & 0.080 & 0.506 & 46.7 & 850.6 & \\
$f_{\mathrm{8}}$ & 31.65313(2) &  & 366.3557 & 0.065 & 0.275 & 38.0 & 580.5 & \\
$f_{\mathrm{9}}$ & 30.19377(2) &  & 349.4650 & 0.064 & 0.562 & 37.2 & 563.3 & \\
$f_{\mathrm{10}}$ & 3.77254(2) & 0.04 & 43.6637 & 0.057 & 0.884 & 31.6 & 446.9 & \\
$f_{\mathrm{11}}$ & 32.35529(2) &  & 374.4825 & 0.055 & 0.554 & 32.4 & 432.9 & \\
$f_{\mathrm{12}}$ & 6.36563(2) & 0.18 & 73.6763 & 0.053 & 0.597 & 29.8 & 395.0 & \\
$f_{\mathrm{13}}$ & 1.15926(3) & -0.05 & 13.4173 & 0.052 & 0.243 & 28.7 & 374.6 & \\
$f_{\mathrm{14}}$ & 1.87679(3) & -0.22 & 21.7221 & 0.044 & 0.401 & 24.7 & 280.3 & \\
$f_{\mathrm{15}}$ & 2.50313(3) & 0.38 & 28.9714 & 0.042 & 0.670 & 23.4 & 253.5 & \\
$f_{\mathrm{16}}$ & 1.51617(5) & 1.11 & 17.5483 & 0.041 & 0.103 & 23.0 & 254.1 & \\
$f_{\mathrm{17}}$ & 29.73612(3) &  & 344.1680 & 0.039 & 0.163 & 22.7 & 217.2 & \\
$f_{\mathrm{18}}$ & 8.75884(4) & 0.71 & 101.3754 & 0.034 & 0.383 & 19.6 & 173.6 & \\
$f_{\mathrm{19}}$ & 3.38343(5) & -0.99 & 39.1601 & 0.029 & 0.698 & 16.0 & 121.7 & \\
$f_{\mathrm{20}}$ & 7.37521(5) & -0.66 & 85.3612 & 0.028 & 0.557 & 15.8 & 115.7 & $f_{\mathrm{3}}+f_{\mathrm{10}}$\\
$f_{\mathrm{21}}$ & 35.37792(5) &  & 409.4667 & 0.025 & 0.664 & 14.8 & 94.6 & \\
$f_{\mathrm{22}}$ & 37.60866(5) &  & 435.2854 & 0.024 & 0.258 & 14.3 & 86.0 & \\
$f_{\mathrm{23}}$ & 11.92258(6) & 0.15 & 137.9929 & 0.022 & 0.674 & 13.0 & 72.6 & \\
$f_{\mathrm{24}}$ & 8.57253(6) & 0.95 & 99.2191 & 0.022 & 0.654 & 12.3 & 70.2 & \\
$f_{\mathrm{25}}$ & 27.04778(6) &  & 313.0530 & 0.021 & 0.033 & 12.1 & 62.9 & \\
$f_{\mathrm{26}}$ & 33.05026(7) &  & 382.5261 & 0.019 & 0.553 & 11.0 & 51.5 & \\
$f_{\mathrm{27}}$ & 3.89026(7) & -0.58 & 45.0261 & 0.019 & 0.298 & 10.5 & 51.2 & \\
$f_{\mathrm{28}}$ & 0.83999(7) & 2.10 & 9.7221 & 0.019 & 0.780 & 10.2 & 49.5 & $f_{\mathrm{2}}-f_{\mathrm{3}}$\\
$f_{\mathrm{29}}$ & 1.67951(7) & -1.71 & 19.4388 & 0.018 & 0.575 & 9.8 & 46.1 & ($2f_{\mathrm{28}}$)\\
$f_{\mathrm{30}}$ & 8.02468(7) & 0.40 & 92.8782 & 0.017 & 0.135 & 9.7 & 43.6 & \\
$f_{\mathrm{31}}$ & 12.39149(8) & -1.47 & 143.4200 & 0.016 & 0.053 & 9.5 & 39.1 & \\
$f_{\mathrm{32}}$ & 5.22971(9) & -0.04 & 60.5291 & 0.015 & 0.166 & 8.4 & 33.3 & \\
$f_{\mathrm{33}}$ & 5.60943(9) & 1.87 & 64.9240 & 0.014 & 0.787 & 7.9 & 29.4 & \\
$f_{\mathrm{34}}$ & 1.56217(9) & -1.43 & 18.0806 & 0.014 & 0.512 & 7.9 & 28.4 & \\
$f_{\mathrm{35}}$ & 14.5764(1) & -0.37 & 168.708 & 0.012 & 0.650 & 7.2 & 22.6 & \\
$f_{\mathrm{36}}$ & 3.6622(1) & -0.95 & 42.387 & 0.012 & 0.135 & 6.9 & 22.1 & $f_{\mathrm{13}}+f_{\mathrm{15}}$\\
$f_{\mathrm{37}}$ & 33.6440(1) &  & 389.398 & 0.012 & 0.977 & 6.9 & 19.9 & \\
$f_{\mathrm{38}}$ & 2.5311(1) & -0.72 & 29.295 & 0.010 & 0.532 & 5.7 & 15.9 & $f_{\mathrm{5}}-f_{\mathrm{13}}$\\
$f_{\mathrm{39}}$ & 7.9278(1) & -1.43 & 91.757 & 0.010 & 0.868 & 5.8 & 16.2 & $f_{\mathrm{12}}+f_{\mathrm{34}}$\\
$f_{\mathrm{40}}$ & 6.6502(1) & 2.25 & 76.970 & 0.010 & 0.425 & 5.8 & 15.5 & \\
$f_{\mathrm{41}}$ & 7.7748(1) & -1.34 & 89.986 & 0.010 & 0.261 & 5.6 & 14.7 & \\
$f_{\mathrm{42}}$ & 35.0501(1) &  & 405.673 & 0.010 & 0.286 & 5.8 & 14.4 & \\
$f_{\mathrm{43}}$ & 1.1769(2) & 0.69 & 13.622 & 0.009 & 0.357 & 5.1 & 13.1 & \\
$f_{\mathrm{44}}$ & 2.1616(1) & 0.65 & 25.018 & 0.009 & 0.397 & 5.1 & 12.7 & $f_{\mathrm{11}}-f_{\mathrm{9}}$\\
$f_{\mathrm{45}}$ & 4.8035(1) & -0.17 & 55.596 & 0.009 & 0.145 & 5.0 & 11.9 & $f_{\mathrm{12}}-f_{\mathrm{34}}$\\
$f_{\mathrm{46}}$ & 1.2225(2) & -1.07 & 14.150 & 0.009 & 0.514 & 4.8 & 10.9 & $f_{\mathrm{32}}-f_{\mathrm{1}}$\\
$f_{\mathrm{47}}$ & 3.3593(2) & 0.75 & 38.881 & 0.008 & 0.586 & 4.6 & 9.9 & ($4f_{\mathrm{28}}$)\\
$f_{\mathrm{48}}$ & 1.1976(2) & 4.00 & 13.861 & 0.008 & 0.381 & 4.5 & 8.0 & \\
$f_{\mathrm{49}}$ & 8.8862(2) & -2.26 & 102.850 & 0.008 & 0.775 & 4.7 & 9.8 & $2f_{\mathrm{2}}$\\
$f_{\mathrm{50}}$ & 34.1421(2) &  & 395.163 & 0.008 & 0.003 & 4.8 & 9.9 & $f_{\mathrm{7}}+f_{\mathrm{34}}$\\
$f_{\mathrm{51}}$ & 1.5106(2) & -5.66 & 17.483 & 0.008 & 0.180 & 4.5 & 8.6 & $f_{\mathrm{6}}-f_{\mathrm{24}}$\\
$f_{\mathrm{52}}$ & 2.5667(2) & 0.99 & 29.707 & 0.008 & 0.992 & 4.4 & 9.2 & $f_{\mathrm{2}}-f_{\mathrm{14}}$\\
$f_{\mathrm{53}}$ & 2.2978(2) & 3.10 & 26.595 & 0.007 & 0.382 & 4.1 & 8.0 & \\
$f_{\mathrm{54}}$ & 31.8377(2) &  & 368.492 & 0.007 & 0.223 & 4.3 & 8.0 & \\
$f_{\mathrm{55}}$ & 1.2767(2) & 0.03 & 14.776 & 0.007 & 0.494 & 4.1 & 7.8 & \\
\hline
\end{tabular}
\tablefoot{
The errors given in parentheses are standard uncertainties derived from the least-squares fitting.
We present the frequencies, amplitudes and phases derived 
from the SC light curve, as the SC data allow the direct detection of peaks above 24.5\,d$^{-1}$.
Furthermore, we list the frequency differences of the SC and LC frequencies in the third column.
The spectral significances (sig) were calculated using \textsc{SigSpec} \citep{2007A&A...467.1353R}.\\
\tablefoottext{a}{$\Delta f_i = (f_{i,\mathrm SC} - f_{i,\mathrm LC}) \times 10^4$ [d$^{-1}$].}
}
\end{table*}

\subsection{Stability investigations}

Strong amplitude variability on month- and year-long timescales was reported in the case of the A--F-type 
hybrid \object{KIC\,8054146} \citep{2012ApJ...759...62B}. We also examined the long-term stability of the most dominant peaks. 
We used the LC data set, because it spans a much longer time interval than the SC data. We investigated the nine 
highest-amplitude frequencies below the LC Nyquist limit. First, we calculated the actual frequencies and their 
amplitudes for each quarter (Q1--Q17). Then, we determined $\chi^2_{\mathrm{red}}$ values that characterise the 
observed variability of these parameters as the deviation of the quarterly data points from their time-averages, relative 
to the uncertainties of the measurements. We did not detect variability in any of the nine frequencies, as 
$\chi^2_{\mathrm{red}} \leq 1.3$ for each of them. 

\begin{figure}
\begin{center}
\includegraphics[width=9.0cm]{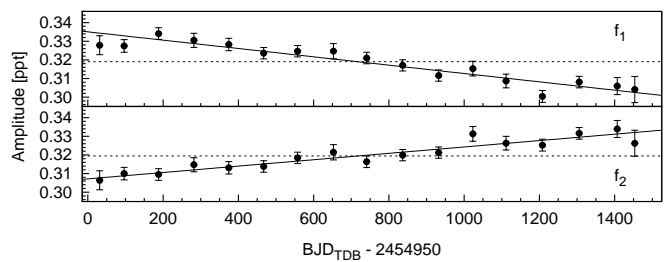}
\end{center}
\caption{Amplitudes of $f_{\mathrm{1}}$ and $f_{\mathrm{2}}$ in the different light-curve segments of the 
Q1--Q17 LC data. The dashed line shows the error-weighted average of the data points, the continuous line 
is a linear fit to the data.}
\label{fig:ampvar}
\end{figure}

However, this test revealed the variability of two amplitudes. 
We obtained $\chi^2_{\mathrm{red}}$ values for the amplitudes of the two most dominant frequencies, 
$f_{\mathrm{1}}$ and $f_{\mathrm{2}}$, of 8.3 and 4.8, respectively. The $\chi^2_{\mathrm{red}}$ value of the other seven 
amplitudes remained below 1.5. Fig.~\ref{fig:ampvar} presents the amplitude values of $f_{\mathrm{1}}$ and $f_{\mathrm{2}}$ 
for Q1--Q17. The amplitudes show a linear trend and vary similarly but in opposite directions. The continuous lines are linear 
fits to the data points, which show a 2.4 and 2.0 per cent relative amplitude variation per year. 
The $\chi^2_{\mathrm{red, lin}}$ values for the linear fits are 1.1 and 0.6, for $f_{\mathrm{1}}$ and $f_{\mathrm{2}}$, 
respectively. These amplitude changes could be explained in terms of energy exchange between the two dominant modes and interpreted as coupling.

\begin{figure}
\begin{center}
\includegraphics[width=9.0cm]{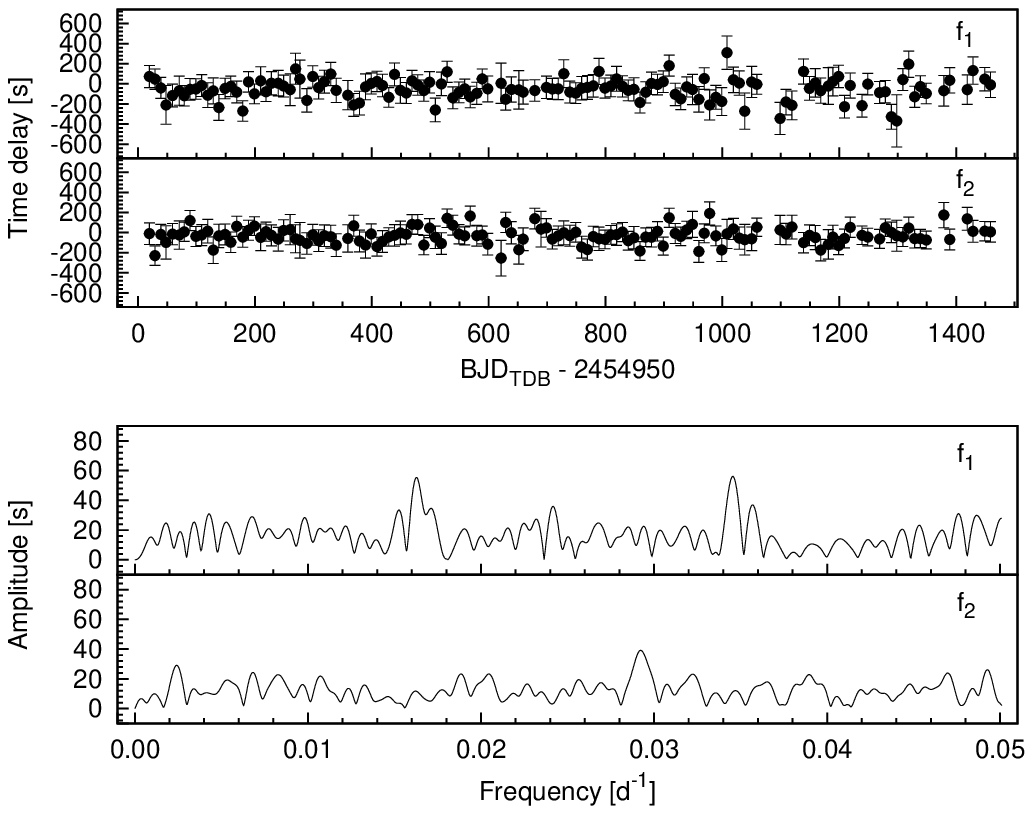}
\end{center}
\caption{Time delays calculated for the two dominant modes $f_{\mathrm{1}}$ and 
$f_{\mathrm{2}}$ (top panels), and their Fourier transform (bottom panels).
}
\label{fig:TD}
\end{figure}

We also checked whether we can detect any sign of phase modulation (PM) of the dominant modes, applying the method
presented by \citet{2014MNRAS.441.2515M}. A periodic phase modulation would indicate the presence of a stellar or even 
a substellar companion of KIC\,9533489. The panels of Fig.~\ref{fig:TD} show the time delays of $f_{\mathrm{1}}$ and 
$f_{\mathrm{2}}$, calculated from 10-days segments of the LC data, and their Fourier transforms, respectively.
The formula used to calculate the time delays is $\tau_{i,j} = (\phi_{i,j} - \overline{\phi_j}) / (2\pi\nu_j)$, 
where $\overline{\phi_j}$ is the mean of the $\phi_j(t)$ phases derived from the different light curve segments for a 
fixed $\nu_j$ frequency \citep{2014MNRAS.441.2515M}.
None of these point to any periodic variation, that is, KIC\,9533489 is not a member of a $20\,\mathrm{d} < P_{\mathrm{orb}} < 4$\,yr binary 
system: neither the time delay plots, nor the corresponding Fourier transforms 
have a common significant peak corresponding to a possible common orbital period.

\subsection{Frequency and period spacings}

In the case of a genuine hybrid, \mbox{KIC\,9533489} will show simultaneously excited high-order $g$- and low-order $p$-modes. 
Thus, we checked, whether
there are any regularities between the periods and frequencies in the presumed $g$- and $p$-mode regions, respectively, 
predicted by stellar pulsation theory.
For references on the theoretical background of non-radial pulsations, 
see e.g. \citet{1980ApJS...43..469T}, \citet{1989nos..book.....U} and \citet{2010aste.book.....A}.

High-radial-order $g$-modes are excited in $\gamma$\,Dor pulsators. According to the first-order asymptotic approximation 
of non-radial stellar pulsations, the separation of consecutive overtones 
having the same $\ell$ spherical harmonic index ($\Delta P_\ell = P_{n+1,\ell} - P_{n,\ell}$) is:

\begin{equation} 
\Delta P_{\ell} \simeq \Pi_0 / \sqrt{\ell(\ell+1)}, 
\label{eq:aps}
\end{equation}
\noindent where
\begin{equation*}
\label{eq:asympeq}
\Pi_0= 2 \pi^2 \left[ \int_{r_1}^{r_2} \frac{N}{r} dr\right]^{-1},
\end{equation*}

\noindent and $N$ is the Brunt-V\"ais\"al\"a frequency. That is, period spacings are expected 
to be constant in a chemically homogeneous star, but inhomogeneities (rapid changes of the mean molecular weight in the 
transition zones) cause deviations from this constant value (see e.g. \citealt{2008MNRAS.386.1487M} 
and \citealt{2011A&A...531A.145B}). 
Following Eq.~\ref{eq:aps}, the period spacing ratios for the different $\ell$ modes are: 

\begin{subequations}
\label{eq:aps2}
\begin{align}
\Delta P_{\ell=1} / \Delta P_{\ell=2} \simeq \sqrt{3} = 1.73, \\
\Delta P_{\ell=2} / \Delta P_{\ell=3} \simeq \sqrt{2} = 1.41.
\end{align}
\end{subequations}

In the case of low-radial-order $p$-modes, even though the pulsation modes are in the non-asymptotic regime for most
of the $\delta$\,Sct stars, we still can notice a clustering of frequencies
with the presumed large separation, $\Delta \nu_\ell = \nu_{n+1,\ell} - \nu_{n,\ell}$ (see e.g. \citealt{2009A&A...506...79G} 
and \citealt{2013A&A...557A..27P}).
The large separation can be calculated in a first approximation as: 
\begin{equation*}
\Delta \nu = 2  \left[ \int_{r_1}^{r_2} \frac{dr}{c_s} \right]^{-1},
\end{equation*}

\noindent where $c_s$ is the sound speed.
The equation given for the large separation (integral over the inverse sound speed) is valid for high radial order $p$-modes 
\citep{1980ApJS...43..469T, 1990LNP...367..283G}. It was generalized in terms of the mean density of the stars by \citet{2014A&A...563A...7S}, based 
on numerical calculations using a huge model grid. The relation obtained for the $\delta$\,Sct stars domain can be approximated by 
\begin{equation}
\label{eq:largesep}
\Delta \nu / \Delta \nu_\odot = 0.776 (\rho / \rho_\odot)^{0.46} 
\end{equation}
\noindent where
$\Delta \nu_\odot = 134.8\,\mu$Hz and $\overline{\rho_\odot} = 1.408$\,g\,cm$^{-3}$.

Another well-known relation in the case of $\delta$\,Sct pulsations 
is the frequency-ratio of the radial fundamental and first-overtone radial mode (see \citealt{1979ApJ...227..935S}): 

\begin{equation}
\label{eq:arany}
f_{\mathrm{FU}}/f_{\mathrm{1O}} \approx 0.77-0.78. 
\end{equation}

In the case of high-overtone ($n\gg\ell$) modes and slow rotation, the frequency differences of the rotationally split components 
can be calculated by the asymptotic relation: 

\begin{equation}
\label{eq:rot}
\delta \nu_{n,\ell,m} = \delta m (1-C_{n,\ell}) \Omega, 
\end{equation}
\noindent where the coefficient \mbox{$C_{n,\ell} \approx 1/\ell(\ell+1)$} for $g$-modes, $C_{n,\ell} \approx 0$ for $p$-modes, 
and $\Omega$ is the (uniform) rotation frequency. As $\delta \nu \sim [1-1/\ell(\ell+1)]$, the ratio of the rotationally 
split $\ell=1$ and $\ell=2$ $g$-modes is in the case of equal $\delta m$:

\begin{equation}
\label{eq:rot2}
\delta \nu_{\ell=1, g} / \delta \nu_{\ell=2, g} \simeq 0.60.
\end{equation}

\noindent We also have in a first approximation:

\begin{equation}
\label{eq:rot3}
\delta \nu_{p} \simeq 2 \delta \nu_{\ell=1, g}.
\end{equation}

\noindent
We investigated the frequency set of KIC\,9533489 considering these relations.

Note that in the case of rapidly rotating stars, additional terms to Eq.~\ref{eq:rot} have to be considered
in the calculation of rotationally split frequencies, considering higher-order rotational effects. In this case,
$\delta \nu_{n,\ell,m}$ also depends on $\nu_{n,\ell}$ and on $m$ ($m_2^2-m_1^2$). For a theoretical
background, see \citet{1990MNRAS.242...25G} and \citet{1998ESASP.418..385K} for the acoustic modes, and \citet{1978AcA....28..441C} 
and \citet{1992ApJ...394..670D} for the $g$-modes.
In Sect.~\ref{sect:freqspacing}, we use the simple form of Eq.~\ref{eq:rot} to derive a first approximation for the rotation
period of the star.

\subsubsection{Frequency ratios}

Considering the modelling results of \citet{2014A&A...563A...7S} on main-sequence $\delta$\,Sct stars, knowing the mean density of 
KIC\,9533489, we estimated the 
theoretical radial fundamental mode's frequency. Taking into account the estimated radius and mass, and their uncertainties (see 
Table~\ref{table:params}), the mean density of the star normalized to the solar value may be between 
0.2 and 0.6\,$\rho/\rho_\odot$. 
These values indicate that the radial fundamental mode could lie between $\sim13-23$\,d$^{-1}$ (150--260\,$\mu$Hz), 
practically in the observed frequency gap, according to fig.~2 of \citet{2014A&A...563A...7S}. 
This means that the radial fundamental mode is either $f_{\mathrm{35}}$ (14.58\,d$^{-1}$), the only mode between 
$13-23$\,d$^{-1}$, or the radial modes are not exceeding the S/N$=4$ significance level.
There is no first overtone frequency pair for $f_{\mathrm{35}}$ with frequency ratio in the range of 0.77--0.78 
according to Eq.~\ref{eq:arany}. 

As the radial fundamental mode represents the low frequency limit for $p$-modes, we can assume that the frequencies
below the gap (up to 14.6\,d$^{-1}$) are `$\gamma$\,Dor-type' $g$-modes. Rapid rotation may explain both the shifts of $g$-modes into higher 
frequencies than the `classical' $\gamma$\,Dor pulsation domain \citep{2013MNRAS.429.2500B} and the lack of radial modes. 
Indeed, rapidly rotating $\delta$\,Sct stars generally tend to show mainly nonradial modes of lower amplitude in comparison with
the slow rotators (see e.g. \citealt{2000ASPC..210....3B}). It is also possible that some of the observed 
frequencies above 5\,d$^{-1}$ are non-zonal ($m\neq0$) modes generated by rotational splittings.

\subsubsection{Frequency spacings and stellar rotation}
\label{sect:freqspacing}

\begin{figure}[ht!]
\begin{center}
\includegraphics[width=8.7cm]{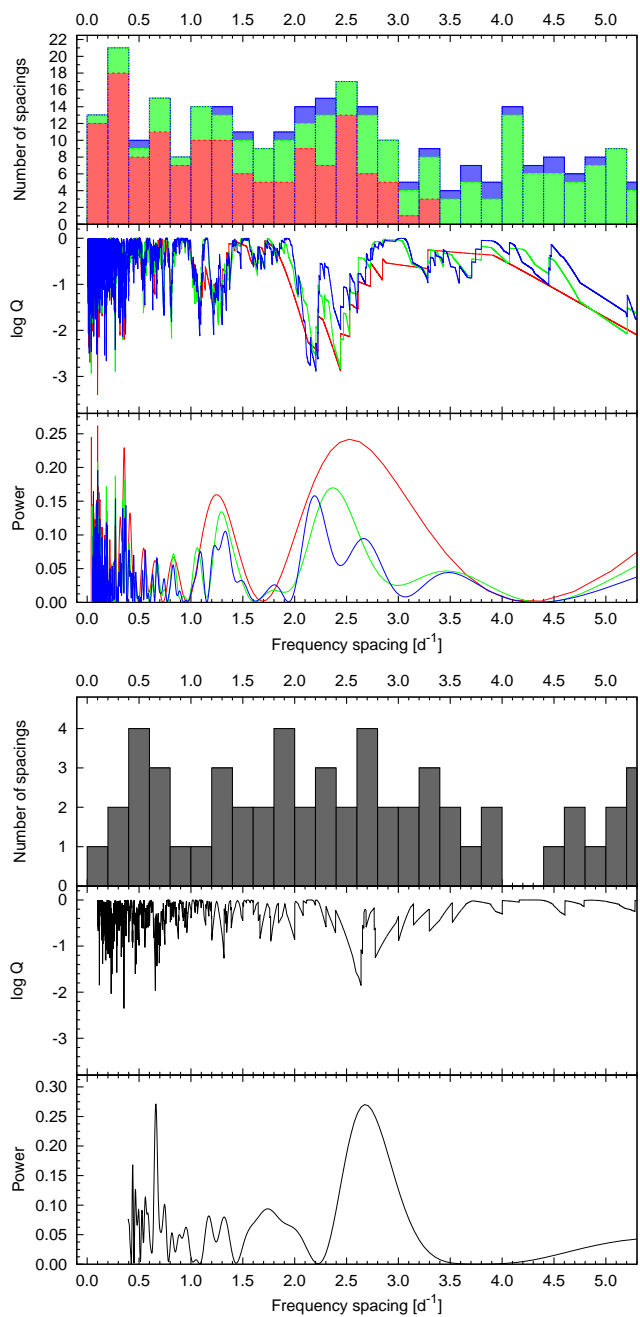}
\end{center}
\caption{Searching for characteristic frequency spacings. \textit{From top to bottom:} histograms of frequency
differences, results of K-S tests, Fourier transforms of different frequency sets. \textit{Red boxes/lines:} 
frequencies below 5\,d$^{-1}$, \textit{green boxes/lines:} frequencies below 10\,d$^{-1}$, \textit{blue boxes/lines:} 
frequencies below 15\,d$^{-1}$, \textit{gray boxes/lines:} frequencies above 27\,d$^{-1}$.
}
\label{fig:spacing}
\end{figure}

We studied the distribution of the frequencies applying different methods: by calculating their Fourier transforms 
(see e.g. \citealt{1997MNRAS.286..303H}), generating histograms of the frequency differences 
(e.g. in \citealt{2007CoAst.150..333M}), and using Kolmogorov-Smirnov (K-S) tests \citep{1988IAUS..123..329K}.
We defined and analysed several sets of frequencies: those below 5, 10, 15, and above 27\,d$^{-1}$.
In the Fourier tests, we assigned equal amplitude to all frequencies. Local maxima in the FTs indicate that characteristic 
spacings exist between the frequencies. In the K-S test, the quantity $Q$ is defined as the probability that the 
observed frequencies are randomly distributed. Thus, any characteristic frequency spacing in the frequency spectrum should appear as 
a local minimum in $Q$. We present the results of these computations in Fig.~\ref{fig:spacing}.

Provided that both $g$- and $p$-modes occur, we expect to find periodicities that belong to rotationally split 
frequencies, according to Eqs.~\ref{eq:rot}, \ref{eq:rot2} and \ref{eq:rot3}. However, note that these relations are 
valid for slow rotators only, and in the case of KIC\,9533489 ($v\,\mathrm{sin}\,i = 70\pm7$\,km\,s$^{-1}$), there might be 
deviations from the constant frequency spacings. Fig.~\ref{fig:spacing} indicates possible characteristic spacings in the 
range of 0.2--0.4, 1.0--1.4 and 2.0--2.7\,d$^{-1}$ in the low-frequency domain, and in the range of 0.4--0.7 and 
2.6--2.7\,d$^{-1}$ above 27\,d$^{-1}$ in the high-frequency domain.

We listed the possible rotational periods of KIC\,9533489, calculated by 
Eq.~\ref{eq:rot} for some specific spacings in Table~\ref{tabl:spac}. These values correspond to local maxima of the FT of the frequencies 
below 10\,d$^{-1}$ (green line in Fig.~\ref{fig:spacing}) or above 27\,d$^{-1}$.
We also considered the possibility that in the case of a high inclination
of the pulsation axis, or if any mode selection mechanism according to $m$ is in operation, we may not detect the $m=0$ 
components of the $\ell=1$ 
triplets, and the spacings belong to $\delta m = 2$ splittings. For completeness, we also listed the possible rotational periods
for $\ell=2$ modes.
We assume that the pulsation and rotation axes coincide.

To check if the two dominant modes ($f_{\mathrm{1}}$ and $f_{\mathrm{2}}$) are rotationally split 
frequencies, we included their 0.44\,d$^{-1}$
separation in Table~\ref{tabl:spac}, too. Finally, we included the case of $f_{\mathrm{34}}$ 
(1.5622\,d$^{-1}$). It takes part in two combinations: $f_{\mathrm{39}} = f_{\mathrm{12}}+f_{\mathrm{34}}$ and 
$f_{\mathrm{45}} = f_{\mathrm{12}}-f_{\mathrm{34}}$. This practically means that there is an equidistant triplet structure
around $f_{\mathrm{12}}$ (6.3656\,d$^{-1}$), with the separation of the value of $f_{\mathrm{34}}$. There are
no other similar structures found, but 1.56\,d$^{-1}$ is another frequency separation we have to take into account as 
the possible rotational spacing.

According to Eq.~\ref{eq:rot3}, we expect that $\delta \nu_{p} \simeq 2 \delta \nu_{\ell=1, g}$. 
Considering the 0.37 and 0.66\,d$^{-1}$ characteristic spacings of the presumed $g$- and $p$-modes, these might correspond 
to rotational splitting values of $\ell=1$ modes. Another possibility is that the 1.30 and 2.67\,d$^{-1}$ spacings are caused by 
the stellar rotation.  

None of these spacings selected in the $g$-mode region shows clearly the 0.6 ratio expected for $\ell=1$ and $2$ modes 
(see Eq.~\ref{eq:rot2}). 

Utilizing the rotation periods calculated from the different spacings, the estimated radius of the star, and the observed 
$v\,\mathrm{sin}\,i$ (see Table~\ref{table:params}), we indicated for guidance the corresponding equatorial velocity ($v$) 
and inclination values in Table~\ref{tabl:spac}. This reveals that since the lower limit for $v$ is $\approx 63$\,km\,s$^{-1}$ 
($v\,\mathrm{sin}\,i = 70-7$\,km\,s$^{-1}$, $i = 90^\circ$, cf. Table~\ref{table:params}), the 0.37\,d$^{-1}$ separations of 
the presumed $g$-modes cannot be interpreted as rotational spacings. This also holds for the 0.66\,d$^{-1}$ spacing of the presumed $p$-modes,
but not for the 0.44\,d$^{-1}$ frequency separation (cf. Table~\ref{tabl:spac}). Therefore 0.44\,d$^{-1}$ is a plausible rotational spacing.

The (critical) Keplerian velocity of $v_{\mathrm {Kep}} = 437$\,km\,s$^{-1}$ is reached at an inclination of $i = 9.4^\circ$.
It seems improbable that the 
2.37\,d$^{-1}$ spacing represents rotationally split $\ell=1, \delta m=1$ $g$-modes, as the derived equatorial velocity of 
371\,km\,s$^{-1}$ is more than 80 per cent of $v_{\mathrm {Kep}}$.

The 2.67\,d$^{-1}$ spacing found from the high-frequency region requires further investigation. We have discussed already that it is
a potential rotational spacing value. However, it can also be related to the large separation of frequencies.
There are known cases in the literature, when such peaks in the Fourier transforms of the observed frequencies 
turned out to be at one half of the large separation (e.g. \citealt{1997MNRAS.286..303H}, 
\citealt{2009A&A...506...79G}). It is expected when modes with different $\ell$ values 
are observed. \citet{2014A&A...563A...7S} present a method to estimate the large separation ($\Delta \nu$) 
for $\delta$\,Sct stars (Eq.~\ref{eq:largesep}). Applying their formula, in the case of KIC\,9533489 
($\rho / \rho_\odot \simeq 0.4$), the 
large separation is expected to be $68.6\,\mu$Hz (5.9\,d$^{-1}$). Half of this value is 2.95\,d$^{-1}$, which is 
not far from the 2.67\,d$^{-1}$ spacing observed. Considering the uncertainties in the radius and mass determination, and 
that equidistant spacings are expected for slow rotators and frequencies in the asymptotic regime, the 2.67\,d$^{-1}$ 
value is a reasonable estimate for half of the large separation.

\begin{figure}
\begin{center}
\includegraphics[width=8.7cm]{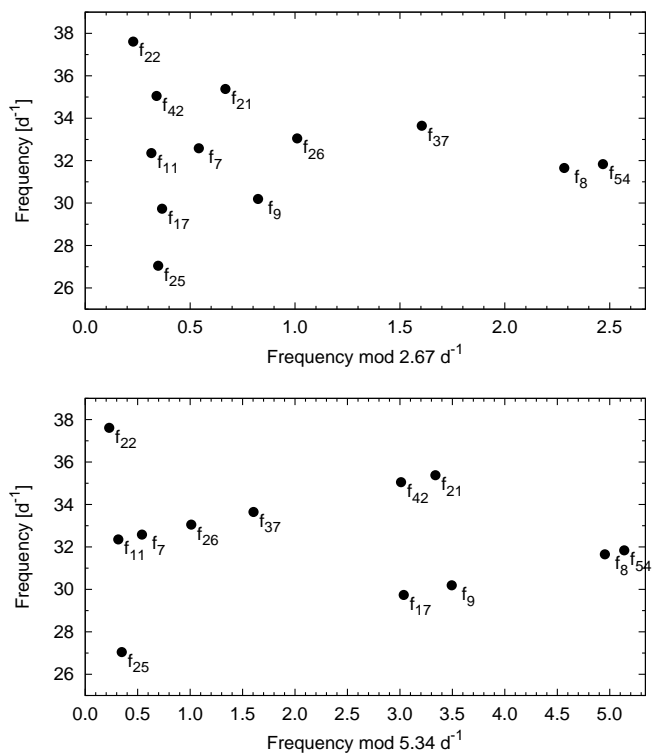}
\end{center}
\caption{Echelle diagrams of the 12 independent frequencies, supposed $p$-modes, above the frequency gap.
}
\label{fig:echelle}
\end{figure}

We demonstrate in Fig.~\ref{fig:echelle} that this 2.67\,d$^{-1}$ frequency spacing is not a mathematical artefact, but
indeed exists in the frequency set. The plot shows echelle diagrams; the frequencies above the gap as a function
of frequency modulo both the 2.67\,d$^{-1}$ spacing and its double. Forming vertical structures, 5-7 frequencies
may be associated with these spacings.

In conclusion of this short discussion of the frequency spacings, we see two options here. One of them is that the pulsator has a high
inclination ($i > 70^\circ$) and is a moderate rotator with an equatorial velocity of around 70\,km\,s$^{-1}$. In this case,
frequency differences with $\approx0.4$\,d$^{-1}$ in the $g$-mode region can be related to rotational splitting.
This also could explain the presence of the 0.84\,d$^{-1}$ peak and its harmonics in the light curve's Fourier
spectrum, as being caused by rotational modulation.

Another option is that \mbox{KIC\,9533489} is a fast rotator, with $v \approx 200$\,\,km\,s$^{-1}$ and $i \approx 20^\circ$. 
This could explain the triplet with 1.56\,d$^{-1}$ frequency separation or the 1.1-1.3\,d$^{-1}$
characteristic spacings shown in Fig.~\ref{fig:spacing} as rotational spacings. In this case, the rotational spacing for 
the $p$-modes and half of their theoretical large separation would lie close to each other.

\begin{table*}
\centering
\caption{\label{tabl:spac} 
Possible rotational periods ($P$), equatorial velocities ($v$) and inclinations ($i$) of KIC\,9533489 applying different 
frequency spacing values.
}
\begin{tabular}{rrrrrrrrrrrrrr}
\hline\hline
 \multicolumn{14}{l}{Low-frequency region (below 10\,d$^{-1}$)} \\
 & & \multicolumn{6}{c}{Dipole modes ($\ell=1$)} & \multicolumn{6}{c}{Quadrupole modes ($\ell=2$)} \\
 \multicolumn{2}{c}{Spacing} & \multicolumn{2}{c}{$P$ [d]} & \multicolumn{2}{c}{$v$ [km\,s$^{-1}$]} & \multicolumn{2}{c}{$i$ [deg]} & 
 \multicolumn{2}{c}{$P$ [d]} & \multicolumn{2}{c}{$v$ [km\,s$^{-1}$]} & \multicolumn{2}{c}{$i$ [deg]}\\
$[$d$^{-1}]$ & [$\mu$Hz] & \multicolumn{1}{c}{$\delta m$=1} & \multicolumn{1}{c}{$\delta m$=2} & \multicolumn{1}{c}{$\delta m$=1}
 & \multicolumn{1}{c}{$\delta m$=2} & \multicolumn{1}{c}{$\delta m$=1} & \multicolumn{1}{c}{$\delta m$=2}
 & \multicolumn{1}{c}{$\delta m$=1} & \multicolumn{1}{c}{$\delta m$=2} & \multicolumn{1}{c}{$\delta m$=1}
 & \multicolumn{1}{c}{$\delta m$=2} & \multicolumn{1}{c}{$\delta m$=1} & \multicolumn{1}{c}{$\delta m$=2}\\
\hline
0.37 & 4.3 & 1.4 & 2.7 & 57.9 & 28.9 & -- & -- & 2.3 & 4.5 & 34.7 & 17.4 & -- & -- \\[1.2mm]
1.30 & 15.0 & 0.4 & 0.8 & 203.4 & 101.7 & 20.1$_{\,18.0}^{\,22.2}$ & 43.5$_{\,38.3}^{\,49.2}$ & 0.6 & 1.3 & 122.0 & 61.0 & 35.0$_{\,31.1}^{\,39.1}$ & -- \\[1.2mm] 
2.37 & 27.4 & 0.2 & 0.4 &  370.7 & 185.4 & 10.9$_{\,9.8}^{\,12.0}$ & 22.2$_{\,19.9}^{\,24.5}$ & 0.4 & 0.7 & 222.4 & 111.2 & 18.3$_{\,16.5}^{\,20.3}$ & 39.0$_{\,34.5}^{\,43.8}$ \\[1.2mm]
0.44 & 5.1 & 1.1 & 2.3 & 68.8 & 34.4 & --$_{\,66.2}^{\,-}$ & -- & 1.9 & 3.8 & 41.3 & 20.6 & -- & -- \\[1.2mm]
1.56 & 18.1 & 0.3 & 0.6 & 244.0 & 122.0 & 16.7$_{\,15.0}^{\,18.4}$ & 35.0$_{\,31.1}^{\,39.1}$ & 0.5 & 1.1 & 146.4 & 73.2 & 28.6$_{\,25.5}^{\,31.7}$ & 73.0$_{\,59.4}^{\,-}$ \\
\multicolumn{5}{l}{High-frequency region (above 27\,d$^{-1}$)} & & & & & & & \\[1.2mm]
0.66 & 7.7 & 1.5 & 3.0 & 51.8 & 25.9 & -- & -- \\[1.2mm]
2.67 & 30.9 & 0.4 & 0.7 & 209.1 & 104.6 & 19.6$_{\,17.5}^{\,21.6}$ & 42.0$_{\,37.1}^{\,47.4}$\\
\hline
\end{tabular}
\tablefoot{
Rotational periods were calculated according to Eq.~\ref{eq:rot}, equatorial velocities
are derived from $R_*$ and $P$, and inclinations from these $v$ values and assuming $v\,\mathrm{sin}\,i = 70\pm7$\,km\,s$^{-1}$.
}
\end{table*}

\subsubsection{Period spacings}

\begin{figure}
\begin{center}
\includegraphics[width=8.7cm]{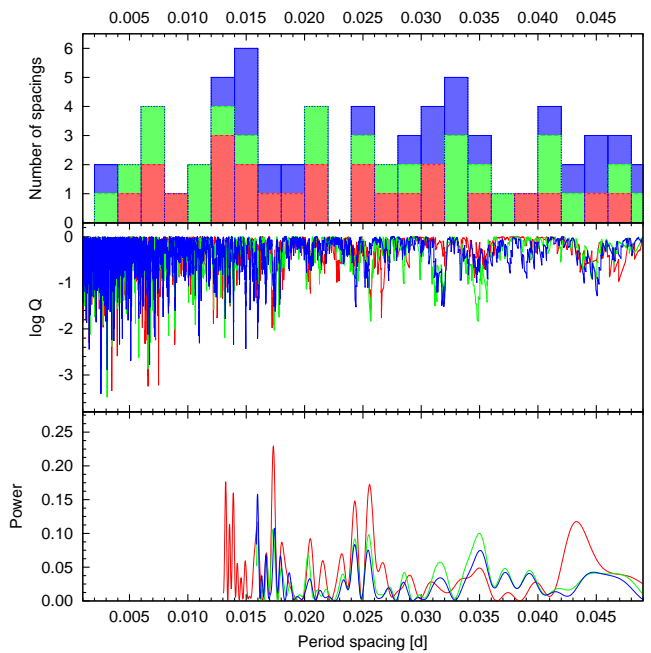}
\end{center}
\caption{Searching for characteristic period spacings. Notations of the different period sets are the same
as in Fig.~\ref{fig:spacing}.
}
\label{fig:spacing_p}
\end{figure}

Using Eqs.~\ref{eq:aps} and \ref{eq:aps2}, we performed similar tests on the periods as on the frequencies in 
Sect.~\ref{sect:freqspacing}. Only the histograms and the
Fourier transforms suggest the presence of some characteristic spacings around 0.012-0.017, 0.002, 0.024-0.026 and 
0.032-0.035\,d (cf. Fig.~\ref{fig:spacing_p}). We cannot use and interpret these indefinite results.
Modelling of KIC\,9533489 could help to decide if any of these spacing values indeed characterise $\ell$ modes 
of the star.

Note that period spacings are strongly influenced not only by chemical gradients, but also by the stellar rotation, especially in the 
case of moderate and fast rotators, as demonstrated by the theoretical study of \citet{2013MNRAS.429.2500B}.

\subsubsection{Other regularities}

Coupling between $g$- and $p$-modes was reported in many hybrids. In the case of \object{\textit{CoRoT}\,105733033}
\citep{2012A&A...540A.117C}, \object{\textit{CoRoT}\,100866999} \citep{2013A&A...556A..87C}, and 
\object{KIC\,11145123} \citep{2014MNRAS.444..102K}, frequencies can be detected in the $p$-mode regime
according to the relation $f_{i,p} = F \pm f_{i,g}$, where $f_{i,g}$ is a $g$-mode frequency and $F$ is
the dominant $p$-mode: the radial fundamental or a low-overtone frequency. The detection of such
coupling supports the idea that the frequencies found in the low- and high-frequency region originate from
the same pulsator.
We note that KIC\,11145123
is a remarkable `textbook example' showing how the regular spacings and their deviations can be used to study 
the stellar interior. Another example of 
mode coupling is the case of \object{KIC\,8054146} \citep{2012ApJ...759...62B}, where peaks were detected 
in the high-frequency region at $f_0 + nf_1$, where $f_0$ is a constant and $f_1$ is the dominant frequency
in the \textit{low-frequency} regime.

We searched for such regularities in the frequency spectrum of KIC\,9533489, but only one frequency in the 
$p$-mode regime can be explained as the combination of another high-frequency peak and a $g$-mode frequency,
as $f_{\mathrm{50}}$ = $f_{\mathrm{34}}+f_{\mathrm{7}}$ (cf. Table~\ref{table:freq}), where 
$f_{\mathrm{50}} = 34.14$, $f_{\mathrm{34}} = 1.56$
and $f_{\mathrm{7}} = 32.58$\,d$^{-1}$. It is worth noting that $f_{\mathrm{34}}$ corresponds to the frequency
separation of a triplet, as discussed in Sect.~\ref{sect:freqspacing}.  

\section{Transit events}
\label{sect:transit}

The light curve of KIC\,9533489 shows transits, which are shallow (0.5\%), short (0.07\,d) and recur every 
197 days (it is also planetary candidate KOI-3783.01). 

First, we checked the field of KIC\,9533489 searching for possible 
external sources, which could be responsible for the observed transit events. There are 
indeed two known stars offset by $\sim 4$ and $8$ arcsec that are blended within the \textit{Kepler} photometric aperture. 
However, these are both too faint to cause 0.5 per cent deep eclipses even if fully eclipsed.
Additionally, the \textit{Kepler} data centroid analysis
for this $40\sigma$ transit event (averaged over all events) shows that the transit host is coincident with the target star
within the $3\sigma$ limit of $0.25\arcsec$.
Considering the adaptive optics observation (see Sect.~\ref{sect:ao}), we know that
there is an additional star $\approx1\arcsec$ from the primary object. However, it appears to be too faint
to be the transit host star, and it is also still too far away from the primary.

We modelled the {\it Kepler} LC and SC data using the {\sc jktebop} 
code (see \citealt{2008MNRAS.386.1644S} and references therein), and accounted for the low sampling rate of the LC data by 
numerically integrating the model to match. Whilst the SC data cover only one transit, their better sampling 
rate makes them more valuable than the LC data, which cover six transits. We pre-whitened the data to remove 
the effects of the pulsations, then extracted the regions around each transit, and normalised them to unit 
flux by fitting a straight line to the out-of-transit data.

\begin{figure}
\centering
\includegraphics[width=7.1cm]{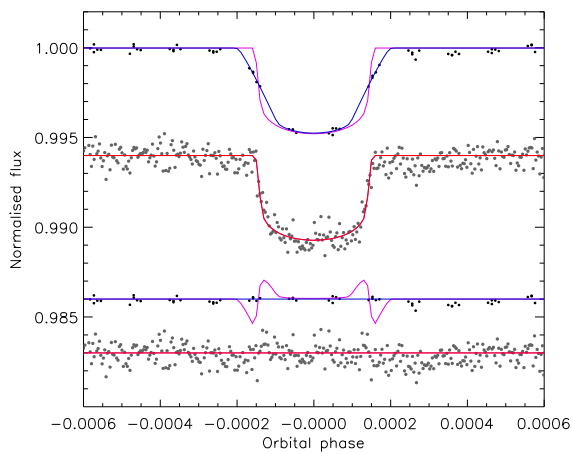}
\caption{\label{fig:withoutL3} {\it Kepler} LC (above) and SC (below)
light curves compared to the best fits from {\sc jktebop} neglecting third light. The residuals are plotted at
the base of the figure, offset from unity. The blue and the purple lines through the LC data
show the best-fitting model with and without numerical integration, and the ones through the residuals show
the difference between this model with and without numerical integration.} 
\end{figure}

The transits are indicative of a planetary-mass object orbiting the F0 star, but we encountered problems with 
this hypothesis. The duration is far too short for this scenario and implies a much smaller radius for the F0 
star than found from our spectroscopic analysis. The low ratio of the radii implied by the transit depth also 
yields transit light curves with ingress and egress durations too short to match the data. These problems forced 
us to consider the possibility that the putative planet is orbiting a fainter star, which is spatially coincident 
with the F0 star. This allows the use of a larger ratio of the radii, yielding a better match to the durations of 
the partial phases of the transit, and also implies a smaller stellar radius and thus a shorter transit (Fig.~\ref{fig:withoutL3}).

\begin{figure}
\centering
\includegraphics[width=7.1cm]{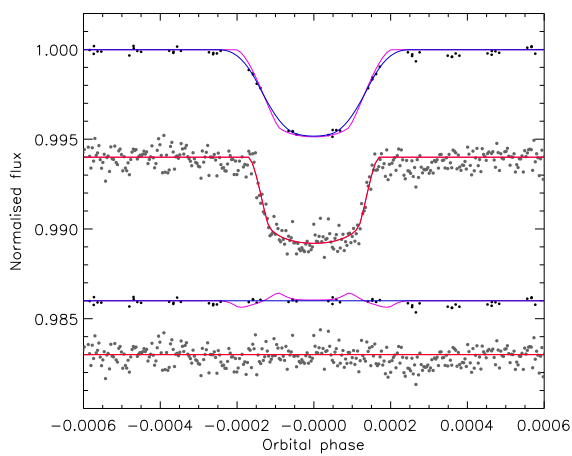}
\caption{\label{fig:withL3} As Fig.~\ref{fig:withoutL3} but allowing for third light when calculating the best fit.} 
\end{figure}

Fitted models of the light curve including a larger `third light' ($L_3$) component are indeed significantly 
better than those with $L_3 = 0$ (see Fig.~\ref{fig:withL3}). \citet{2012MNRAS.426.1291S} showed that $L_3$ is very poorly 
constrained when modelling 
transits in isolation, but only considered values between 0.0 and 0.75 (expressed as a fraction of the total light 
of the system). We find that $L_3 = 0.88 \pm 0.03$ yields a determinate and improved fit to the light curve. 
The fainter star should therefore be roughly eight times fainter than the F0 star, which implies a mass of about 
$0.95\,M_\odot$ if the two stars are at the same distance. We used the DSEP theoretical stellar evolutionary models 
\citep{2008ApJS..178...89D} to guide these inferences of the physical properties of the system.

Whilst a $0.95\,M_\odot$ star has a much smaller radius than the F0 star, we still require a large orbital 
eccentricity for the transit model to match the observations: $e\sin\omega = 0.956$ where $e$ is orbital 
eccentricity and $\omega$ is the argument of periastron. The large $L_3$ value we find means that a larger 
ratio of the radii ($k=0.185$ versus $0.064$) is needed to match the transit depth. Such an object orbiting a 
$0.95\,M_\odot$ main sequence star will have a radius of approximately $0.16\,R_\odot$ ($1.6\,R_{\rm Jup}$). 
This is consistent with the radius of an inflated transiting planet (see \citealt{2012MNRAS.426.1338S}), but also with that 
of a low-mass star. We therefore cannot claim that the transiting object is a planet.

If we allow the transit host star to be closer to Earth than the F0 star, making them merely an asterism rather 
than gravitationally bound, then its contribution to the system light can be matched by a less massive and thus 
smaller star. This would mean that the light curve could be matched using a smaller $e\sin\omega$, and also 
imply a lower radius for the transiting object. We do not currently have sufficient observational constraints 
to constrain the relative distances of the two stars. Table~\ref{tab:sysprop} gives a set of plausible physical 
properties of the system, based on the transit light curves, the DSEP theoretical stellar models, the assumption 
that the F0 star and the transit host star have a common distance, and that the transiting body is stellar rather 
than an inflated exoplanet.

\begin{table}
\centering
\caption{\label{tab:sysprop} Plausible model of the KIC\,9533489 system.
}
\begin{tabular}{l l l l} 
\hline
\hline
\multicolumn{4}{c}{Orbital parameters} \\
$T_0$ (BJD$_{\mathrm {TDB}}$)         & \multicolumn{3}{l}{2455527.46780 $\pm$ 0.00025} \\
$P$ (d)                 & \multicolumn{3}{l}{197.14468 $\pm$ 0.00015} \\
$e\sin\omega$           & 0.956           & & \\
$e\cos\omega$           & 0.0 fixed       & & \\
\hline
\multicolumn{4}{c}{Photometric parameters} \\
$r_{\rm A}= R_{\rm A} / a$             & 0.0060 fixed    & $i$ ($^\circ$)          & 89.97\\
$k = R_{\rm B}/R_{\rm A}$                     & 0.185           & $L_3$                   & 0.875\\
$r_{\rm B}= R_{\rm B} / a$             & 0.0011          & & \\
\hline
\multicolumn{2}{r}{Transit host star} & \multicolumn{2}{l}{transiting body} \\
Mass ($M_\odot$)            & 0.95            & 0.15 & \\
Radius ($R_\odot$)          & 0.85            & 0.16 & \\
$T_{\mathrm{eff}}$ (K)              & 5500            & 3200 & \\
\hline
\end{tabular}
\tablefoot{$L_3$ is a fraction of the total light 
in the {\it Kepler} passband which is \textit{not} coming from the transit host star, $a$ is the orbital semimajor 
axis and $R_{\rm A,B}$ are the true radii of the objects.
}
\end{table}

\section{Conclusions}

KIC\,9533489 is one of the many A--F spectral type stars observed by \textit{Kepler}.
It was selected as a hybrid candidate pulsator by \citet{2011A&A...534A.125U}. What makes it a promising target
for detailed analysis is its hybrid nature which is most probably genuine. Besides that, it also shows periodic transit-like events
in its light curve.

One of the relevant questions is whether KIC\,9533489 is a genuine hybrid showing simultaneously self-excited 
$p$- and $g$-modes or not. The asteroseismic relevance of such genuine hybrids is very high, as their structure 
can be probed down to the deep interior layers. Based on the spectroscopically derived physical parameters 
of KIC\,9533489 showing its location within the overlapping region of the $\delta$\,Sct and $\gamma$\,Dor instability strips, 
on the dominant modes' coupled amplitude variation, and on the strong multiperiodicity in the low 
frequency domain, we argue that KIC\,9533489 is a genuine hybrid pulsator. Furthermore, based on frequency spacing investigations,
we come to the conclusion that KIC\,9533489 is either a highly inclined moderate rotator 
($v\approx 70$\,km\,s$^{-1}$, $i > 70^\circ$) or a fast rotator with $v\approx 200$\,km\,s$^{-1}$ and 
$i\approx20^\circ$. We need a comprehensive modelling of the star which, among other parameters, takes into account the 
effects of rotation.

The transit analysis revealed that KIC\,9533489 (or the KIC\,9533489 system) is unusual.
It seems, that we have to consider the possibility of a high-order system:
the \textit{Kepler} aperture contains (1) an F0 star showing pulsations, (2) a fainter star at $1\arcsec$
(detected by the adaptive optics measurements), (3) an (unseen) G/K-type star in the AO PSF of the F0 star 
(either gravitationally bound or unbound), and (4) a small/faint object transiting the G/K star. Since stellar 
binaries (or higher order systems) are common, it is certainly possible to have a hierarchical triple 
system (objects 1, 3, 4). Further radial velocity observations on long (years) time scale
and higher signal-to-noise ratio spectra might help to define the spectral type of the unseen transit
host star and thus help to characterise the transiting object, too.

\begin{acknowledgements}
The authors thank the anonymous referee for the constructive comments and recommendations on the manuscript.
Zs.B. acknowledges the support of the Hungarian E\"otv\"os Fellowship (2013), the kind hospitality of 
the Royal Observatory of Belgium as a temporary voluntary researcher (2013--2014), and, together with \'A.S.,
the support of the ESA PECS project 4000103541/11/NL/KML.
\'A.S. acknowledges support by the Belgian Federal Science Policy (project M0/33/029, PI: P.D.C.) and 
by the J\'anos Bolyai Research Scholarship of the Hungarian Academy of Sciences.
P.M.F. acknowledges support from MICINN of Spain via grants BES-2012-053246 and AYA2011-24728, and from the 
``Junta de Andaluc\' ia''  through the FQM-108 project.
Funding for the \textit{Kepler} mission is provided by the NASA Science Mission directorate. 
We thank the \textit{Kepler} team for the high-quality data obtained by this outstanding mission.
We are very grateful to Jorge Jim\'enez Vicente (Universidad de Granada, Spain) and Evencio Mediavilla (Instituto de Astrof\' isica 
de Canarias) for kindly providing a spectrum with the \textsc{integral} optical fiber system.
Based on spectra obtained with the \textsc{hermes} spectrograph installed at the Mercator Telescope, operated by 
the Flemish Community, with the Nordic Optical Telescope, operated by the Nordic Optical Telescope 
Scientific Association, and with the William Herschel Telescope, operated by the Isaac Newton Group of Telescopes.
All instruments are located at the Observatorio del Roque de los Muchachos, La Palma, Spain, of the Instituto de 
Astrofisica de Canarias.
Some of the data presented herein were obtained at the W.~M. Keck Observatory, which is operated as a scientific 
partnership among the California Institute of Technology, the University of California and the National Aeronautics 
and Space Administration. The Observatory was made possible by the generous financial support of the W.~M. Keck Foundation.
The authors thank Simon Murphy for his useful comments on the first version of this manuscript.

\end{acknowledgements}



\bibliographystyle{aa}
\bibliography{bibliography.bib}


\listofobjects

\end{document}